# Comment on: "Depolarization corrections to the coercive field in thin-film ferroelectrics"

Stephen Ducharme[1] and Vladimir Fridkin[1,2]

[1] Department of Physics and Astronomy, Center for materials Research and Analysis, University of Nebraska, Lincoln, NE 68588-0111, USA

[2] Institute of Crystallography, Russian Academy of Sciences, Moscow 153025, RUSSIA

Email:  sducharme1@unl.edu and fridkin@ns.crys.ras.ru

## Abstract

The Letter by Dawber et al. [J. Phys.:  Condens. Matter **15** L393 (2003)] notes that incomplete screening in the electrodes of a ferroelectric capacitor can result in an underestimate for the true coercive field in films of nanometer thickness. We show that their estimate of the magnitude of this correction it too large in the case of ferroelectric copolymer Langmuir-Blodgett films and, as a result, invalidates the claim that finite-size scaling of the ferroelectric coercive field is evident in films thinner than 15 nm.

The recent letter by Dawber et al. [1] proposes a model of thin-film ferroelectric capacitors accounting for the potential drop across real metal electrodes due to the finite density of carriers available for screening. In the case of ferroelectric switching, the potential drop in the electrodes opposes the internal field of the ferroelectric film and therefore reduces the external potential necessary to switch the polarization. The model concerns an ideal ferroelectric film, with an internal spontaneous polarization $P_S$ and dielectric constant $\varepsilon_f$, bounded by identical electrodes of an ideal metal, with exponential Thomas-Fermi (TF) screening length $\lambda$ and dielectric constant $\varepsilon_e$. This model was used to recalculate coercive field data from ferroelectric copolymer Langmuir-Blodgett (LB) films ranging in thickness from 0.9 nm to 60 nm. [2] The recalculated values were used to underline the universality of an empirical finite-size scaling behavior of the coercive field with the –2/3 power of the film thickness, which behavior is reported for many of ferroelectric materials.

To account for finite screening in the electrodes, the coercive field values were recalculated by Dawber et al. using the expression

$$E_C = \frac{V_C + 2P_S \lambda/\varepsilon_e\varepsilon_0}{d + 2\lambda\varepsilon_f/\varepsilon_e}, \tag{1}$$

where $V_C$ is the minimum voltage required to switch the polarization of a film of thickness $d$, $\varepsilon_0$ is the permittivity of free space and the factors of 2 account for the two electrodes bounding the ferroelectric film. The data in question consists of the values of the coercive field reported as $E_C = V_C/d$ for a series of vinylidene fluoride copolymer LB film ranging in thickness from 0.9 nm to 60 nm, bounded by aluminum electrodes. [2] The material parameters chosen for the recalculation of $E_C$ were $P_S = 0.2$ C/m$^2$ for the ferroelectric film, and $\lambda = 0.045$ nm and $\varepsilon_e = 1$ for the electrodes. [1] With these parameters, the TF contribution represented by the second term of the numerator in Eq. 1 is 2.2 V, or 1.1 V for each electrode, and the 'corrected' coercive field for the thinnest (0.9 nm) LB film is 2.7 GV/m, more than five times the nominal value $V_C/d = 0.5$ GV/m. While the values of $V_C/d$ are independent of thickness for films of 15 nm and thinner, [2] the recalculated $E_C(d)$ data appears to follow the $d^{-2/3}$ power law. [1] In other words, the dependence of $E_C$ on thickness arises from the recalculation using Eq. 1 and is not evident in the original data.

There are several problems with the recalculation using Eq. 1. The measured values of the spontaneous polarization in vinylidene fluoride copolymers [3] range from 0.05 to 0.1 C/m$^2$, no more than half the value used by Dawber et al. The dc dielectric constant of the screened portion of the electrode $\varepsilon_e$ must be larger than unity because of contributions from the Fermi surface and from the atomic core polarizabilities. Also, the second term in the denominator of Eq. 1 was neglected in the analysis, but it equals 0.7 nm and is not negligible compared to the thickness of the thinnest films. Sot the choice of parameters results in an overestimate of the values of the coercive field, especially for the thinnest LB films. Further, the Thomas-Fermi exponential screening form used in the derivation of Eq. 1 is only the lowest approximation ("zeroeth order," according to Ziman [4]), as it is derived from the independent free-electron gas model, thus ignoring the Fermi surface and electron-electron interactions. The typical estimated screening length of ~1/2 Angstrom for metals amounts to an internal contradiction because the atomic potentials are far from smooth at this length scale. [5] Many-body calculations show that the exponential TF form used to derive Eq. 1 is a poor approximation, even in the bulk, [6] and the interface presents additional complications. [7] Therefore, Eq. 1 is adequate only for estimates of the electrode potentials, even with accurate values of the material parameters.

Our more recent measurements of the dielectric properties of the ferroelectric copolymer LB films place an upper limit on the ratio $\lambda/\varepsilon_e$ as follows. We made a series of LB-film capacitors from the same copolymer used in the coercive field study, [2] a random copolymer consisting of 70% vinylidene fluoride and 30% trifluoroethylene, by methods described in detail previously. [8] The films ranged in thickness from 1 to 120 monolayers (ML) and were bounded top and bottom by crossed strip electrodes of aluminum such that each film consisted of up to 6 independent capacitors. Capacitance measurements were made with an HP 4292B Impedance Analyzer operating at 1 kHz with a 0.1 V amplitude. Figure 1 shows that the reciprocal specific capacitance $A/C$ of these films is proportional to the number of layers. [9] The data are consistent with a simple linear dependence of the form with slope 0.0203±0.0004 mm$^2$/nF/ML and intercept –0.02±.02 mm$^2$/nF. The low-amplitude ac measurement is sensitive only to the denominator of Eq. 1, where $C/A = (d/\varepsilon_f\varepsilon_e + 2\lambda/\varepsilon_e\varepsilon_0)$. According to Eq. 1, there should be an intercept of $2\lambda/\varepsilon_e\varepsilon_0$ = +0.01 mm$^2$/nF, which is inconsistent with the data. Therefore, the electrode potentials are much smaller than estimated by Dawber et al.

In conclusion, we have shown that the model of Dawber et al. is only approximate, that their application of the model overestimates the potential drop in the electrodes, and that the data from ferroelectric copolymer LB films do not exhibit finite-size scaling of the coercive field below 15 nm.

## Acknowledgments

We thanks Alexander Sorokin for providing the data for Figure 1. This work was supported by the National Science Foundation, the Nebraska Research Initiative, and the Russian National Science Foundation.

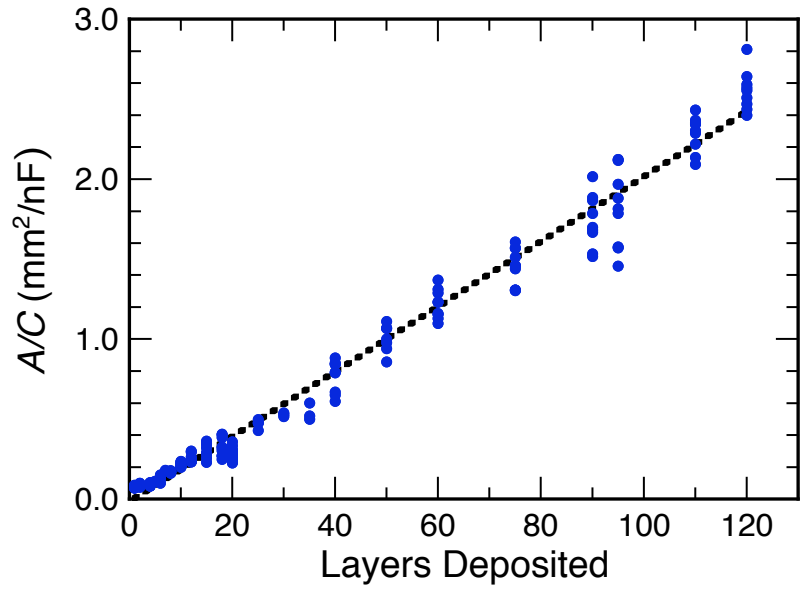

**Figure 1:** Reciprocal specific capacitance for a series of copolymer LB film capacitors. The dotted line is a linear regression fit to all the data.